\documentclass[journal]{IEEEtran}
\pdfoutput=1 
\usepackage{graphicx} 
\usepackage{amsmath, amssymb} 
\usepackage{algorithm, algpseudocode} 

\usepackage[hidelinks]{hyperref} 
\usepackage{setspace} 
\usepackage[font=small]{caption} 
\usepackage[utf8]{inputenc} 
\usepackage{geometry} 

\IEEEoverridecommandlockouts 

\title{Aero-engines Anomaly Detection using an Unsupervised Fisher Autoencoder}

\author{Saba Sanami and Amir G. Aghdam
\thanks{Saba Sanami and Amir G. Aghdam are with the Department of Electrical and Computer Engineering, Concordia University, Montreal, QC, Canada. Email: {\tt\small saba.sanami@concordia.ca, amir.aghdam@concordia.ca}}
\thanks{This work has been supported by the Natural Sciences and Engineering Research Council of Canada (NSERC) under grant RGPIN-2024-06367}}

\begin{document}

\maketitle
\thispagestyle{empty}
\pagestyle{empty}

\begin{abstract}

Reliable aero-engine anomaly detection is crucial for ensuring aircraft safety and operational efficiency. This research explores the application of the Fisher autoencoder as an unsupervised deep learning method for detecting anomalies in aero-engine multivariate sensor data, using a Gaussian mixture as the prior distribution of the latent space. The proposed method aims to minimize the Fisher divergence between the true and the modeled data distribution in order to train an autoencoder that can capture the normal patterns of aero-engine behavior. The Fisher divergence is robust to model uncertainty, meaning it can handle noisy or incomplete data. The Fisher autoencoder also has well-defined latent space regions, which makes it more generalizable and regularized for various types of aero-engines as well as facilitates diagnostic purposes. The proposed approach improves the accuracy of anomaly detection and reduces false alarms. Simulations using the CMAPSS dataset demonstrate the model's efficacy in achieving timely anomaly detection, even in the case of an unbalanced dataset.

\end{abstract}

\section{Introduction}

Detecting anomalies in aero-engines is essential for mitigating potential risks, maintaining consistent operational reliability, and enhancing overall safety. Any deviation from the normal operation of an aero-engines may have severe consequences, including compromised performance or even catastrophic failure. An anomaly detection algorithm acts as an early warning system, enabling engineers and maintenance personnel to identify irregularities or potential issues in engine function before they cause serious problems. These detection systems provide timely insights into engine impairment, facilitating proactive maintenance measures and minimizing the risk of unexpected flight failures \cite{imp1,imp2,imp3,imp4}.

Existing methods for aero-engine anomaly detection can be broadly classified into three categories: model-based, data-driven, and hybrid methods. Model-based methods use physical models or mathematical equations to describe the normal behavior of aero-engines and detect anomalies when the deviation of the output from its expected nominal value exceeds a prescribed threshold value. These methods require a deep understanding of the engine’s structure and dynamics and often fall short in capturing complex (often nonlinear) relationships among engine components \cite{model1}. On the other hand, data-driven methods use statistical or machine learning techniques to learn the normal patterns of aero-engine data and identify faults based on prior knowledge of the engine’s physics. These methods can handle high-dimensional and noisy data, but may suffer from the lack of labeled data, the difficulty of feature engineering, and the challenge of interpreting the results. Hybrid methods combine model-based and data-driven methods to utilize the advantages of both approaches. These methods can incorporate domain knowledge and data analysis to improve the accuracy and robustness of anomaly detection. However, they may also inherit the limitations of both methods, such as the complexity of model building and the dependency on data quality \cite{model2,model3}.

One can use deep learning approaches to overcome the challenges of traditional methods in aero-engines anomaly detection. Unlike conventional methods, deep learning techniques do not require capturing intricate patterns or understanding of nonlinear relationships within complex aero-engines. Instead, these models excel at autonomously learning such patterns from extensive multisensory data, making them highly adaptable to the dynamic nature of aero-engine systems. Since unsupervised deep learning models do not require labeled data, no data annotation problems exist. These advantages, coupled with the advent of the Internet of Things (IoT) and Industry 4.0, along with improved accessibility to high-quality engine data, establish a strong foundation for using deep learning approaches for aero-engine health monitoring. These approaches address challenges inherent in traditional methods while harnessing the advancements in technology, thereby paving the path for enhanced anomaly detection in aero-engine systems \cite{deeplearning1}, \cite{deeplearning2}, \cite{deeplearning3}.

Various studies explore the application of deep learning methods in anomaly detection. The authors in \cite{deep1} introduce a two-step framework consisting of a stacked denoising autoencoder (AE) for extracting features from the input data and a single Gaussian model construction for anomaly detection. Another study proposes a sequential variational AE (VAE) combined with a convolutional neural network (SeqVAE-CNN) for unsupervised multivariate time-series anomaly detection \cite{deep3}. A new approach to detect anomalies in time series using a VAE-LSTM hybrid model was proposed in \cite{deep4}. The model utilizes a VAE module for forming robust local features over short windows and an LSTM module for estimating the long-term correlation in the series on top of the features inferred from the VAE module. Finally, a deep learning model, anomaly VAE-Transformer, which combines the VAE to extract local information in the short term and the transformer to identify dependencies between the data in the long term, was proposed in \cite{deep5}. These studies highlight the advantages of AE and VAE in detecting subtle deviations and complex patterns in aero-engine data.

In this work, we extend the application of Fisher AE (FAE), introduced in \cite{FAE}, to the aero-engine anomaly detection scope. The adaptation of FAE to this context is challenging due to the distinct characteristics of aero-engine data, which include high variability, strong correlations among multivariate time series, and significant noise levels. Moreover, our model incorporates a mixture of Gaussian distributions with learnable parameters for the latent space prior distribution, enabling dynamic adaptation to the data observed. This prior assumes that the data consists of separate clusters in the latent space, and it has been demonstrated to outperform other approaches on challenging datasets \cite{prior}.

By introducing the Fisher criterion into the AE, the model enhances the separation between classes in the latent space while simultaneously minimizing the distance within each class, which can significantly improve the overall effectiveness of the anomaly detection mechanism. One notable advantage of this type of AE is its enhanced ability to reach an interpretable and connected regularized latent space. The maintenance personnel can more quickly and efficiently perform diagnostics, facilitating a clearer understanding of the learned representations. The Fisher information helps the model to focus on capturing features that contribute significantly to the discrimination between normal and anomalous instances. This is crucial in imbalanced datasets where anomalies are less frequent, as the model needs to identify subtle patterns that differentiate the minority class. Additionally, the model's capability to detect anomalies promptly enables proactive maintenance interventions and minimizes potential downtime. Furthermore, the FAE reduces false positives and unnecessary alerts, making the diagnostic process more efficient.  

The rest of the paper is organized as follows. Section~II presents some background and motivation for the study. The FAE is developed and implemented in Section~III, as the main contribution of this work. Comparative simulation results are presented in Section~IV to underscore the model's superiority. Finally, some concluding remarks are given in Section~V.

\section{Problem Statement}

\textbf{Notation:} Throughout the paper, $\mathbb{R}$ and $\mathbb{N}$ represent the sets of real and natural numbers, respectively. The symbol $\|.\|$ denotes the Euclidean norm (also known as the L2 norm), $\nabla_a$ denotes the gradient operator with respect to the variable $a$, and $\Delta_{a}f$ indicates the Laplacian of the function $f$ with respect to $a$.

Aero-engine data acquisition systems record a wide range of time-series data from sensors for evaluating engine performance. These recorded data can be categorized into two groups: static and dynamic. Static measurements encompass performance parameters related to the thermodynamic system, while dynamic variables capture oscillations and vibrations within the engine components during operation. Both types of measurements are vital for fault detection in aero-engines. However, this study focuses on static measurements of turbofan engines. 

The sensors that capture static data may vary for different engines, reflecting the uniqueness of their design and specifications. However, certain variable dynamics are the same for distinct turbofan engine types. These include variables such as the fan inlet temperature and pressure, the physical fan speed, the engine pressure ratio (P50/P2), the ratio of fuel flow to Ps30, the burner fuel-air ratio, and the turbine cool air flow \cite{sensor3}. For a complete list of the engine-independent variables, see \cite{sensor3}.

We aim to utilize these data to design a robust and regularized unsupervised model for detecting anomalies in multivariate data and address the challenges of (i) imbalanced data distribution in aero-engines, (ii) insufficient generalization across various engine models and configurations, and (iii) computational time constraints.

Consider an aero-engine with \( N \in \mathbb{N} \) sensors. Let \( x_i \in \mathbb{R}^D, i \in \{1, \ldots, N\}\),  $D \in \mathbb{N}$,  be the measurements of the \( i \)-th sensor. Define the aggregate sensor matrix $X = [{x_1}, x_2, \ldots, x_N] \in \mathbb{R}^{D \times N}$, and denote the latent variable by $z$. The size of $z$ is a design choice based on the desired level of data compression and the complexity of the data patterns. The problem is then formulated as finding the instances at which
\begin{equation}
\ \|X - f_\theta(z)\|^2 >\lambda.
\end{equation}
Here, $f_\theta(z)$ is the reconstruction of $X$ from $z$, and $\lambda$ is the anomaly threshold. To identify the anomaly points, we propose the robust generative FAE, conceptualized by minimizing the Fisher divergence between the intractable joint distribution of observed data and latent variables, aligning it with the modeled joint distribution. This approach will be discussed in detail in the following section.
\section{Main Results}
A common approach in data fitting and probabilistic analysis involves choosing a probability distribution $p_{\theta}$, that reduces a specific divergence measure in comparison to the unknown true data distribution $p_{\text{true}}$. One well-known distance measure used in this context is the Kullback-Leibler (KL) divergence. By considering the probabilistic model of the observed data as $p_\theta(X)$, the KL divergence is defined as follows
\begin{align*}
\int_{\mathcal{X}} p_{\text{true}}(X) \log\frac{p_{\text{true}}(X)}{p_{\theta}(X)} \, dX,
\end{align*}
where $\mathcal{X}$ is the set of all possible data points $X$. While minimizing KL divergence is useful, it may have limitations when handling complex data. An alternative method called Fisher divergence has been recently introduced as follows \cite{Fisher}
\begin{align}
D_{F}[p_{\text{true}}\|p_{\theta}] = &\int_{\mathcal{X}} p_{\text{true}}(X) \frac{1}{2} \|\nabla_X \log p_{\text{true}}(X) \nonumber \\
&- \nabla_X \log p_{\theta}(X)\|^2 \, dX \nonumber \\
= &\mathbb{E}_{p_{\text{true}}(X)}\left[\frac{1}{2} \|\nabla_X \log p_{\text{true}}(X) \right. \nonumber \\
&\left. - \nabla_X \log p_{\theta}(X)\|^2\right].
\end{align}
Utilizing the results of \cite{Fisher},
\begin{align}\label{Ha}
D_{F} [p_{\text{true}}||p_{\theta}] &= \mathbb{E}_{p_{\text{true}}(x)} \frac{1}{2} \|\nabla_X \log p_{\text{true}}(X)\|^2  + s_{H} [p_{\theta}(X)],
\end{align}
where
\begin{align*}
s_{H} [p_{\theta}(X)] &= \frac{1}{2} \|\nabla_X \log p_{\theta}(X)\|^2 + \Delta_X \log p_{\theta}(X).
\end{align*}

The Fisher metric evaluates how well the model distribution can match the true distribution's structure by comparing their gradients. The measure is more focused on the shape or geometry of the distributions rather than their absolute positions, making it less sensitive to outliers or areas of low probability mass that can disproportionately affect the KL divergence.

Let \(p_{\eta, \theta}(X, z)\) be the joint distribution of the modeled data and corresponding latent variables, and \(q_{\text{true},{\phi}}(X, z)\) the joint distribution of the true observed data and latent variables, where \(\phi\), \(\theta\) and \(\eta\) represent the parameters of the encoder, decoder and the prior distribution of the latent space, respectively. Using the probability chain rule, both joint distributions are defined as follows \cite{FAE}
\begin{align}
q_{\text{true},\phi}(X, z) &= p_{\text{true}}(X)q_{\phi}(z|X) \nonumber, \\
p_{\eta,\theta}(X, z) &= p_{\eta}(z)p_{\theta}(X|z),\label{eq:jointDistributions}
\end{align}
where $q_\phi(z|X)$ is the probability density function of the encoded distribution, also known as the approximate posterior. It describes how the latent variable $z$ is distributed, given the input data $X$. A Gaussian model is often used for the posterior. $p_\theta(X|z)$ is the likelihood of reconstruction, which represents the probability of reconstructing the input data $X$, given the latent space $z$. Moreover, $ p_\eta(z)$ is the prior probability density function of the encoded distribution, representing our assumptions about the distribution of the latent variables before observing any data. In this study,  the prior function is modeled as a mixture of Gaussian distributions with learnable parameters of the means, variances and mixture weights.

It is desired to minimize the Fisher divergence between two joint distributions
\begin{equation} \label{lossfunction}
\begin{aligned}
\phi^*, \eta^*, \theta^* &= \text{argmin}_{\phi, \eta, \theta} {D_{F}} \left[ q_{\text{true},\phi}(X, z) \,\middle\|\, p_{\eta, \theta}(X, z) \right], \\
&=\text{argmin}_{\phi, \eta, \theta}  \mathbb{E}_{q_{\text{true},\phi}(X, z)}[\frac{1}{2} \| \nabla_{X,z} \log q_{\text{true},\phi}(X,z) \\&- \nabla_{X,z} \log p_{\eta,\theta}(X,z) \|^2].
\end{aligned}
\end{equation}
By substituting (\ref{eq:jointDistributions}) into (\ref{lossfunction}) and utilizing (\ref{Ha}), it results from Theorem~1 of \cite{FAE} that the minimization of (\ref{lossfunction}) is equivalent to the following optimization problem
\begin{align}
\phi^*, \eta^*, \theta^* = \arg\min_{\phi,\eta,\theta} & \Bigg\{ \mathbb{E}_{p_{\text{true}}(X)} \Big[ D_{F} \big[ q_{\phi}(z|X) \parallel p_{\eta,\theta}(z|X) \big] \Big] \nonumber \\
& + \mathbb{E}_{q_{\phi}(z|X)} \Big[ s_{H} \big[ p_{\theta}(z|X) \big] \nonumber \\
& + \frac{1}{2} \big\| \nabla_{X} \log q_{\phi,\text{true}}(X|z) \big\|^2 \Big] \Bigg\}.
\end{align}

Considering $p_{\theta}(X|z)$ as Gaussian distribution with mean $f_{\theta}(z)$ and variance 1, and using Monte Carlo estimation with $L$ samples from $q_{\phi}(z|X)$, the loss function to minimize (\ref{lossfunction}) is defined in  \cite{FAE} and \cite{FAE2} as

\begin{equation} \label{lossfunction2}
\begin{aligned}
\mathcal{L} (X; \phi, \eta, \theta) = &\frac{1}{2L}\sum_{l=1}^{{L}} \| \nabla_z \log q_\phi(z^{(l)}|X) \\& - \nabla_z \log p_\eta(z^{(l)}) - \nabla_z \log p_\theta(x|z^{(l)})\|^2 \\
&+ \frac{1}{2} \|X - f_\theta(z^{(l)})\|^2 \\&+ \frac{1}{2} \left\| \nabla_X \log q_\phi(z^{(l)}|X) \right\|^2,
\end{aligned}
\end{equation}
where 
$z^{(l)}$ is the latent variable sampled from $q_\phi(z^{(l)}|x)$ using the reparameterization trick 
$z^{(l)} = \mu(X) + \sigma(X) \odot \epsilon^{(l)}$. Here, $\mu(X)$ and $\sigma(X)$ are the mean and standard deviation predicted by the model, and $\epsilon^{(l)}$ is a random variable with a standard normal distribution $\mathcal{N}(0, I)$. This reparameterization enables one to use backpropagation during the training process. Moreover, $f_\theta(z^{(l)})$ is the reconstruction of $X$ from $z^{(l)}$ using the decoder and $k$ is the regularization control constant.

The loss function in (\ref{lossfunction2}) comprises three components. The first term measures the divergence between the approximate posterior $q_\phi(z^{(l)}|X)$ and the model posterior $p_\theta(X|z^{(l)})$ using the Fisher information matrix of the prior $p_\eta(z^{(l)})$. Anomalies typically cause changes in the data distribution that deviate from the norm, and the Fisher divergence term helps the model adapt to these variations. By minimizing the first term in (\ref{lossfunction2}), the model captures the uncertainty or variability in the data distribution, and becomes more robust to anomalies. The second term, which is the reconstruction error, measures the dissimilarity between the original input data $X$ and its reconstruction $f_\theta(z^{(l)})$. Faults in aero-engine time-series data will likely result in significant deviations from the expected or normal data patterns. By minimizing this term, the model aims to capture and reproduce normal data patterns accurately. The third term is stability control, which penalizes large gradients of the encoder network’s output for its parameters. By minimizing this term, the model extracts stable and invariant features from the input data, making it less sensitive to minor fluctuations and noise.

The anomaly detection process with the FAE begins by training the model on a dataset representing normal conditions. During this phase, the Adam optimizer is utilized to optimize the model parameters. Once the training is complete, the FAE model reconstructs data samples. For each sample, we calculate the reconstruction error. Anomalies are identified by comparing these errors against a predefined threshold, which can be set using statistical methods or based on specific domain knowledge of the data. Samples whose reconstruction error exceeds this threshold are classified as anomalies. This method effectively detects unusual patterns, showcasing the model's capability to identify departures from the established normal behavior. The detailed procedure is outlined in Algorithm~\ref{alg:training_vae} and~\ref{alg:anomaly_detection} .

\begin{algorithm}
\caption{Training the VAE with Fisher Divergence}
\label{alg:training_vae}
\begin{algorithmic}[1]
\State \textbf{Input:} Training data $X$, latent dimension
\State \textbf{Initialize} VAE parameters: $\phi$, $\eta$, $\theta$
\Repeat
    \State Randomly sample a minibatch of training data: $\{X_i\}_{i=1}^N$
    \State Compute the gradient: $\nabla_{\phi, \eta, \theta} \mathcal{L} (X_i; \phi, \eta, \theta)$
    \State Update $\phi$, $\eta$, and $\theta$ using the Adam optimizer
\Until{Convergence }
\State \textbf{Output:} Optimized parameters $\phi^*$, $\eta^*$, $\theta^*$
\end{algorithmic}
\end{algorithm}

\begin{algorithm}
\caption{Anomaly Detection using the Trained VAE}
\label{alg:anomaly_detection}
\begin{algorithmic}[1]
\State \textbf{Input:} Test data $X'$, trained parameters $\phi^*$, $\eta^*$, $\theta^*$, threshold $\lambda$
\State Reconstruct the data $X'_{\text{recon}} = f_{\theta^*}(z)$ where $z = g_{\phi^*}(X')$
\State Calculate reconstruction error $\epsilon = \|X' - X'_{\text{recon}}\|^2$
\State \textbf{if} $\epsilon > \lambda$
    \State \quad \textbf{then} Flag as anomaly
    \State \textbf{endif}

\State \textbf{Output:} Anomaly indicators
\end{algorithmic}
\end{algorithm}

\if {0}
\subsection{Model Generalization}
As discussed earlier, employing the Fisher divergence has the advantage of dealing with probability distributions, increasing the model's robustness. Using Fisher divergence ensures the model is sensitive to anomalies and resilient to false positives, a common challenge in anomaly detection systems. Moreover, the algorithm's performance under varying conditions and with different engine types underscores its adaptability and generalizability, making it a versatile tool in aero-engine maintenance and monitoring.

Overfitting is a common concern in all anomaly detection models. However, incorporating regularization terms in the loss function plays a pivotal role in mitigating the risk of overfitting. The Fisher divergence term, as expressed in (\ref{lossfunction2}), is a regularization mechanism that penalizes deviations from the prior distribution of latent variables. This prevents the model from fitting the training data too closely and promotes a more generalized representation of the latent space. Simultaneously, the stability control term further contributes to regularization by penalizing large gradients of the encoder network's output. This encourages the extraction of stable and invariant features, making the model less susceptible to minor fluctuations and noise in the input data. The combined effect of these terms ensures that the anomaly detection method is not overly sensitive to the noise of the training data, leading to a more robust model that generalizes well to unseen data. The regularization techniques employed in the proposed method effectively balance model complexity, preventing the learning of overly specific features and enhancing the model's adaptability to diverse conditions.

\fi
\section{Simulations}
To evaluate the model's performance, we apply the FAE to the commercial modular aero propulsion system simulation (CMAPSS) dataset. This public dataset, synthetically generated by NASA, comprises simulated run-to-failure data from turbofan engines. The dataset consists of multiple multivariate time series, each of which corresponds to a different engine. Each time series includes sensor readings recorded over time. The dataset includes data fleet of FD001, FD002, FD003, and FD004, each representing different operating conditions and fault modes. These variants provide diverse scenarios for algorithm testing \cite{cmapss}.

\textbf{Example~1}. In the first example, we focus on the FD001 subset of the CMAPSS dataset, which presents scenarios of fault occurring in the high-pressure compressor (HPC). We utilize all 21 measurements in the dataset, including various sensor readings and engine operational conditions. Our approach involves implementing two distinct AEs: VAE and FAE, to detect anomalies in the time series data.

The initial step involves data preprocessing, where we adopt a data augmentation strategy to optimize the training of our deep learning models. This strategy entails duplicating a specific dataset segment that represents normal operational conditions. The primary purpose of this duplication is to increase the representation of normal states within the dataset (in real-world scenarios, this step might be unnecessary), ensuring that the models are adequately trained with typical operational patterns.

\begin{figure}
\centering
	\includegraphics[scale=0.5]{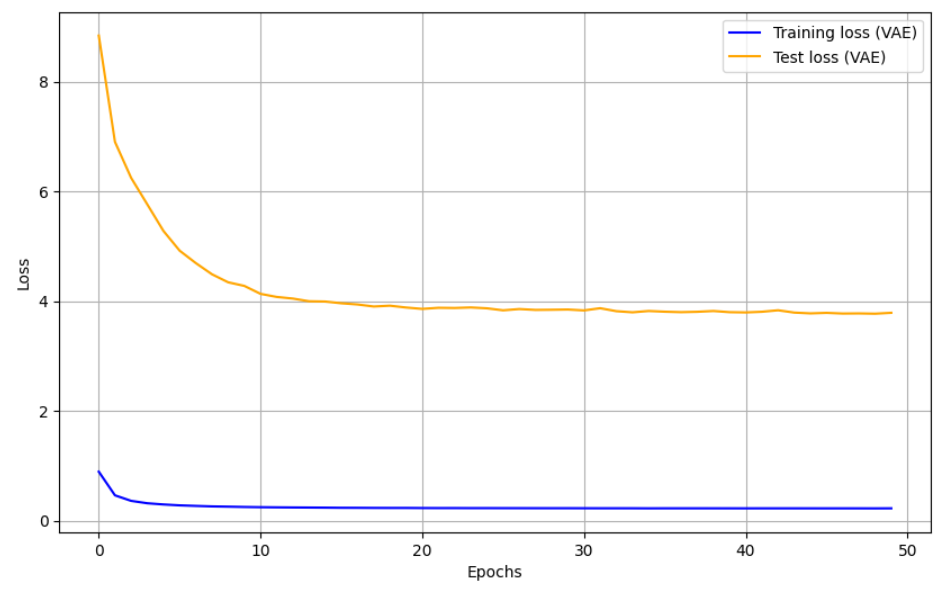}
	\caption{Train and test loss using VAE in FD001 dataset }
	\label{VAE1}
\end{figure}

\begin{figure}
\centering
	\includegraphics[scale=0.5]{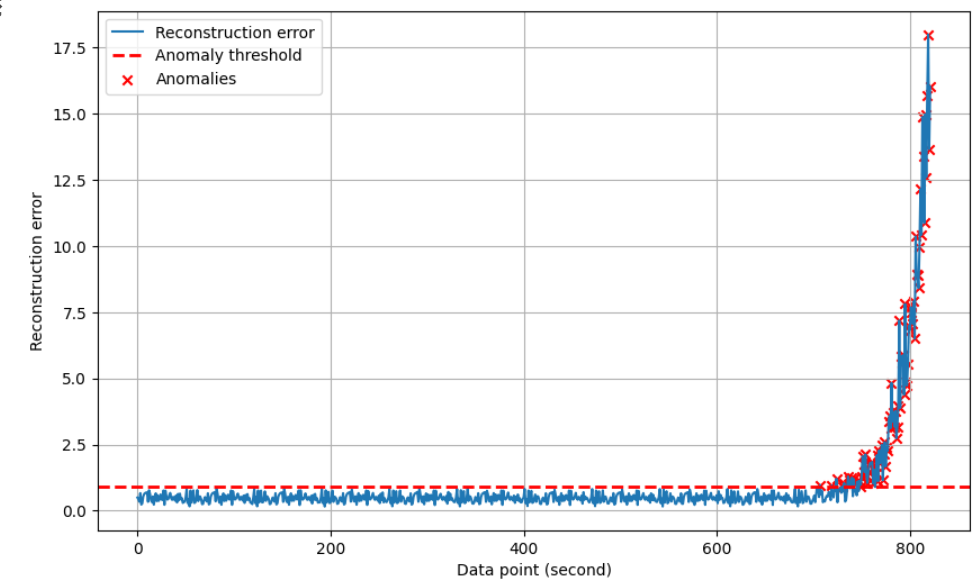}
	\caption{Anomaly detection using VAE in FD001 dataset }
	\label{VAE1AD}
\end{figure}

\begin{figure}
\centering
	\includegraphics[scale=0.5]{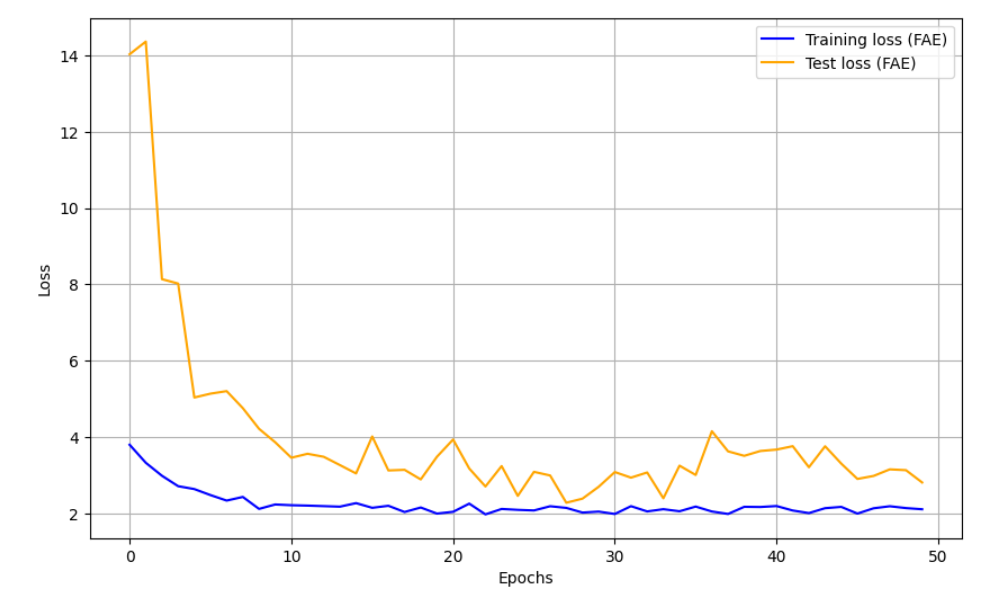}
	\caption{Train and test loss using FAE in FD001 dataset }
	\label{FAELOSS}
\end{figure}

\begin{figure}
\centering
	\includegraphics[scale=0.5]{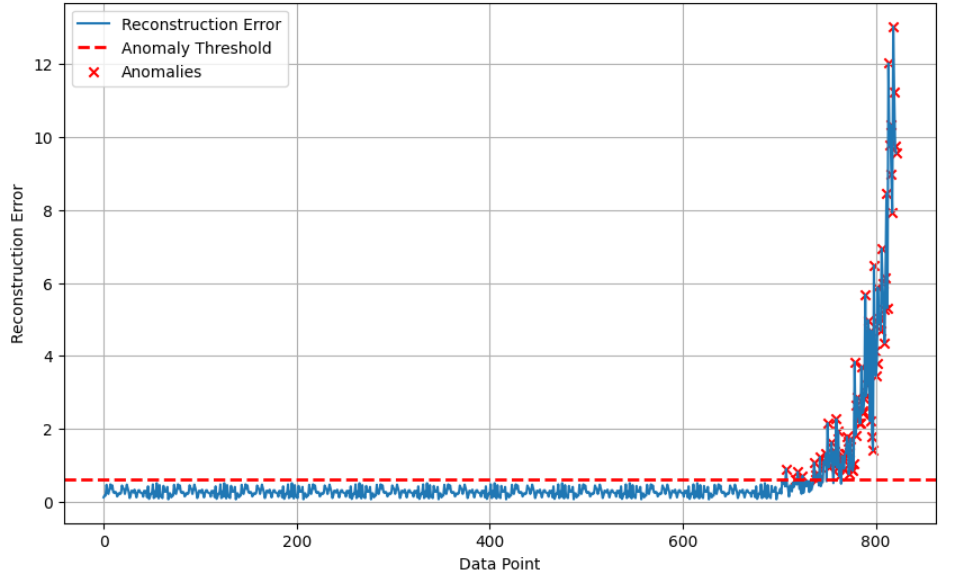}
	\caption{Anomaly detection using FAE in FD001 dataset }
	\label{FAE1AD}
\end{figure}

In the next step, the VAE and FAE are trained with pre-specified hyperparameters in Table \ref{tab:1}. The prior is a mixture of three Gaussian distributions, each with its own set of learnable parameters of the mean and log-variance. Throughout the learning process, these parameters are updated to better capture the complex nature of the data distribution. The model is used to reconstruct the data, calculate reconstruction errors, and identify anomalies by setting a suitable threshold. The threshold is set at the 90th percentile of calculated reconstruction errors. The resulting loss values and reconstruction errors are, respectively, illustrated in Figs.~\ref{VAE1} and \ref{VAE1AD} for the VAE, and Figs.~\ref{FAELOSS} and \ref{FAE1AD}  for the FAE. It can be observed that the FAE is less susceptible to overfitting and has a lower test loss than the VAE. Furthermore, the reconstruction error threshold established by the FAE model effectively reduces the incidences of false positives.

\begin{figure}
\centering
	\includegraphics[scale=0.5]{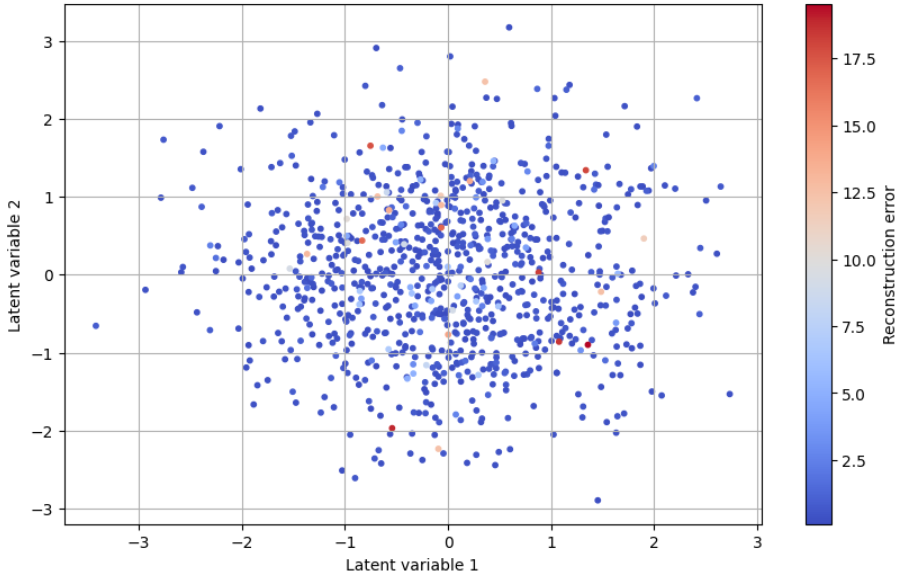}
	\caption{Distribution of latent variables in VAE}
	\label{VAEL2}
\end{figure}

\begin{figure}
\centering
	\includegraphics[scale=0.4]{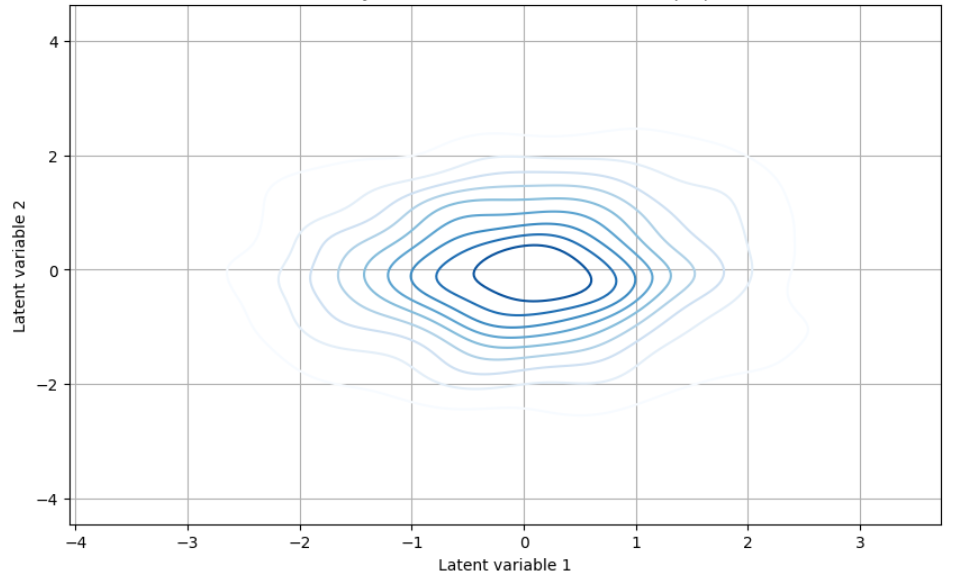}
	\caption{Distribution of latent variables in VAE}
	\label{VAEL1}
\end{figure}

\begin{figure}
\centering
	\includegraphics[scale=0.5]{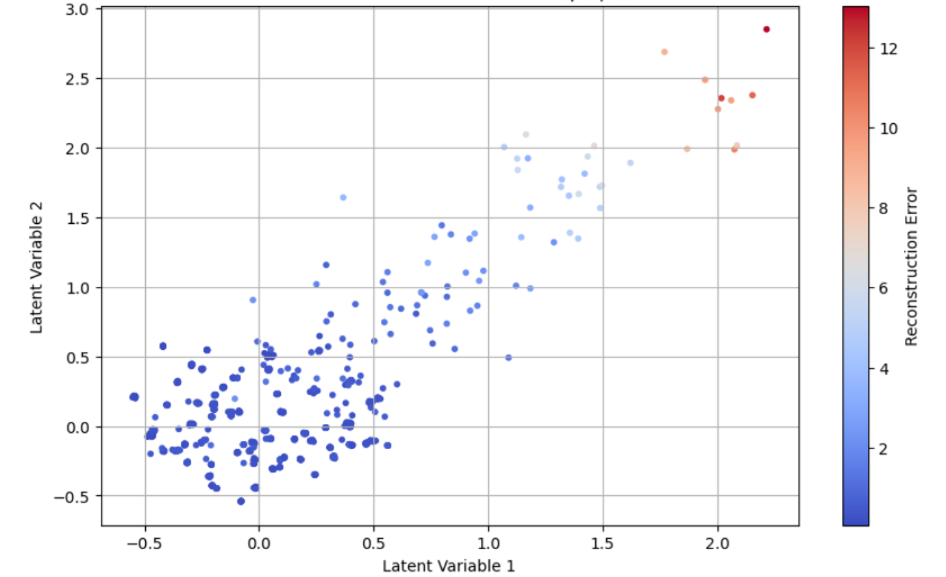}
	\caption{Probability distribution of latent variables in FAE}
	\label{FAEL1}
\end{figure}

\begin{figure}
\centering
	\includegraphics[scale=0.5]{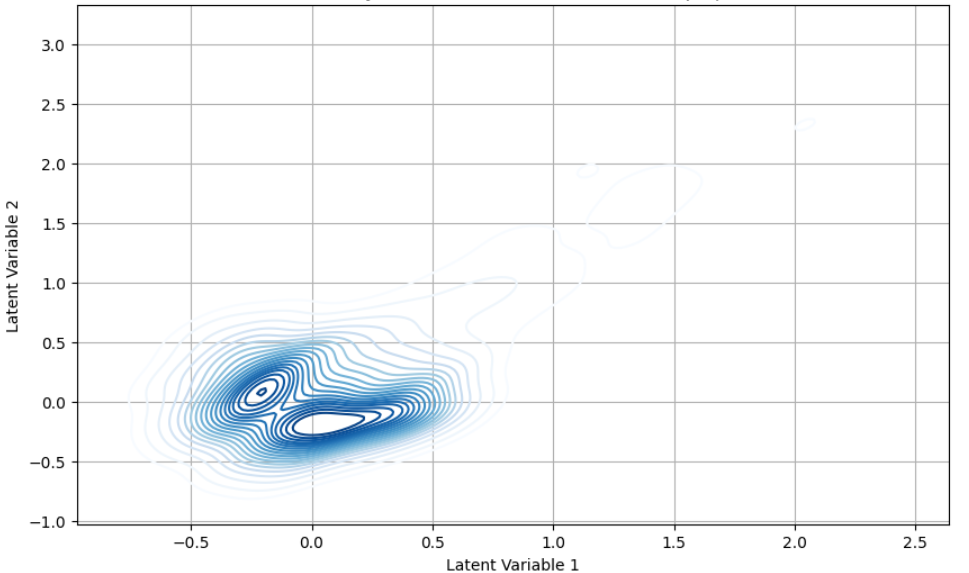}
	\caption{Probability distribution of latent variables in FAE}
	\label{FAEL2}
\end{figure}

Latent space analysis using scatter plots and Kernel density estimation (KDE) plots are presented in Figs.~\ref{VAEL2} and \ref{VAEL1} for the VAE, and in Figs.~\ref{FAEL1} and \ref{FAEL2} for the FAE. These figures offer a deeper understanding of the distribution and separation of normal and anomalous data in the latent space. Comparing the outcomes of both models, it is evident that the FAE provides a more structured and regularized latent space, offering a clearer distinction between normal and abnormal data, as well as more meaningful and interpretable representations of the input data. This demarcation significantly improves the robustness of anomaly detection, allowing for more precise identification of outliers, hence enhancing the FAE's ability to detect anomalies. In contrast, the VAE model does not clearly distinguish between normal and abnormal data, leading to potentially less effective and less reliable results.\\
\begin{table}
    \centering
    \caption{The AEs hyperparameters}
    \begin{tabular}{|c|c|c|c|}
        \hline
        Hyperparameters & Value or description \\
        \hline  
        Encoder hidden layer size & 32, ReLU activation\\
        Decoder hidden layer size & 32, ReLU activation\\
        Latent size & 2\\
        Number of mixture components & 3\\
        Initializer for mixture weights & Uniform\\
        Initializer for mixture means & Zeros\\
        Initializer for mixture log variances & Zeros\\
        Batch size & 16\\
        No. of epochs  & 50\\

        \hline
    \end{tabular}
    \label{tab:1}
\end{table}
\textit{Remark~1}. The parameters of the mixture model can learn to capture various operational states of the aero-engine corresponding to different operating regimes, such as idling, takeoff, and cruising. This capability helps differentiate between normal operational variations and anomalies.\\
\textit{Remark~2}. Leveraging the structured and separate latent space in the FAE approach minimizes the necessity for a decoder layer. This streamlined process contributes to reducing the diagnostic time of the VAE to identify anomalies.

\begin{figure}
\centering
	\includegraphics[scale=0.5]{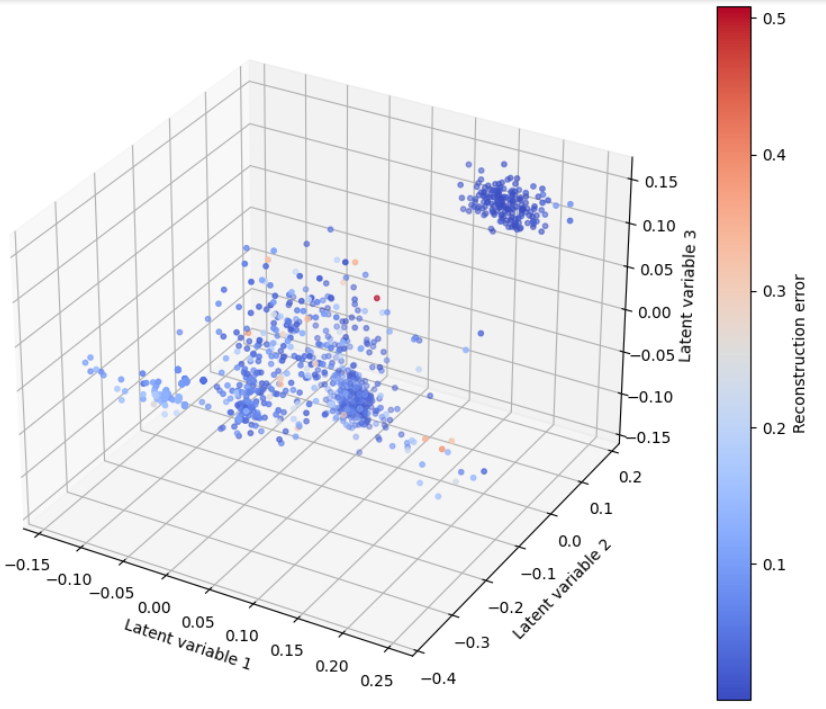}
	\caption{Distribution of latent variables in FAE}
	\label{FAEL4}
\end{figure}

\textbf{Example~2}. In the second example, we apply the FAE to the FD003 subset, where faults originate from the fan and HPC degradation. To address the complexity, we increase the FAE latent variables from two (in Example~1) to three. The distribution of these latent spaces is illustrated in Fig.~\ref{FAEL4}. A clear observation from this figure is the distinct boundary that the FAE establishes between normal and abnormal data. 

\textit{Remark~3}. The latent space structure in Example~2 is noticeably different from that observed in Example~1. This difference underscores the adaptability and diagnostic capabilities of the FAE. The model can detect anomalies and represent different types of faults through specific changes in the configuration of the latent space. This unique characteristic of the FAE, where the latent space structure dynamically adapts to different types of faults, reveals its effectiveness in diagnostic applications. The ability to visually and quantitatively distinguish between various fault sources in the latent space can be pivotal for high-performance diagnosis and targeted maintenance strategies.



\section{Conclusions}

This research investigates the application of the FAE in aero-engine anomaly detection. The approach aims to minimize the discrepancy between the actual and modeled data distributions, allowing the FAE to categorize the aero-engine's data properly. The FAE's ability to create a regularized and well-structured latent space makes it particularly effective in distinguishing between normal and abnormal data patterns, thereby enhancing the precision of anomaly detection. This distinction is crucial for the early identification of potential issues in aero-engines, which is paramount for ensuring flight safety and operational efficiency. Simulations performed on the CMAPSS dataset confirm its efficacy in aero-engine health monitoring. In future work, we plan to explore the model's uncertainty quantification to objectively evaluate risks associated with potential anomalies.



\begin{thebibliography}{00}

\bibitem{imp1} C. Zhang, L. Cui, Q. Zhang, Y. Jin, X. Han, and Y. Shi, ``Online Anomaly Detection for Aeroengine Gas Path Based on Piecewise Linear Representation and Support Vector Data Description," \textit{IEEE Sensors Journal,} vol. 22, pp. 22808-22816, 2022.

\bibitem{imp2} S. Chen, M. Wu, P. Wen, F. Xu, S. Wang, and S. Zhao, ``A Multimode Anomaly Detection Method Based on OC-ELM for Aircraft Engine System," \textit{IEEE Access}, vol. 9, pp. 28842-28855, 2021.


\bibitem{imp3} H. Wang, W. Jiang, X. Deng, and J. Geng, ``A New Method for Fault Detection of Aero-engine Based on Isolation Forest,"\textit{ Measurement}, vol. 185, p. 110064, 2021.

\bibitem{imp4} C. Chen, N. Lu, B. Jiang, Y. Xing, and ZH. Zhu, ``Prediction Interval Estimation of Aeroengine Remaining Useful Life Based on Bidirectional Long Short-term Memory Network," \textit{IEEE Transactions on Instrumentation and Measurement,} vol. 70, pp. 1-13, 2021.


\bibitem{model1} Z. Lei, G. Wen, S. Dong, X. Huang, H. Zhou, Z. Zhang, and X. Chen, ``An Intelligent Fault Diagnosis Method Based on Domain Adaptation and Its Application for Bearings Under Polytropic Working Conditions," \textit{IEEE Transactions on Instrumentation and Measurement}, vol. 70, pp. 1-14, 2020.

\bibitem{model2} Z. Wei, S. Zhang, S. Jafari, and T. Nikolaidis. ``Gas Turbine Aero-engines Real-time On-board Modelling: A Review, Research Challenges, and Exploring the Future," \textit{Progress in Aerospace Sciences}, vol. 121, art. 100693, 2020.

\bibitem{model3} N. Rath, R.K. Mishra, and A. Kushari, ``Aero-engine Health Monitoring, Diagnostics and Prognostics for Condition-based Maintenance: An Overview," \textit{International Journal of Turbo and Jet-Engines}, 2022.

\bibitem{deeplearning1} K. Choi, J. Yi, C. Park, and S. Yoon, ``Deep Learning for Anomaly Detection in Time-series Data: Review, Analysis, and Guidelines, \textit{IEEE Access}, vol. 9, pp. 120043-120065, 2021.

\bibitem{deeplearning2} R. Zhao, R. Yan, Z. Chen, K. Mao, P. Wang, and R. X. Gao, ``Deep Learning and Its Applications to Machine Health Monitoring," \textit{Mechanical Systems and Signal Processing}, vol. 115, pp. 213-237, 2019.

\bibitem{deeplearning3} M. Carratù, V. Gallo, SD. Iacono, P. Sommella, A. Bartolini, F. Grasso, L. Ciani, and G. Patrizi. ``A Novel Methodology for Unsupervised Anomaly Detection in Industrial Electrical Systems," \textit{IEEE Transactions on Instrumentation and Measurement}, 2023.

\bibitem{deep1} X. FU, H. Chen, G. Zhang, and T. Tao, ''A New Point Anomaly Detection Method About Aero-engine Based on Deep Learning," \textit{International Conference on Sensing, Diagnostics, Prognostics, and Control (SDPC)}, pp. 176-181, 2018.


\bibitem{deep3} T. Choi, D. Lee, Y. Jung, and H. Choi, ``Multivariate Time-series Anomaly Detection Using SeqVAE-CNN Hybrid Model," \textit{International Conference on Information Networking (ICOIN)}, pp. 250-253, 2022.

\bibitem{deep4} S. Lin, R. Clark, R. Birke, S. Schönborn, N. Trigoni, and S. Roberts, ``Anomaly Detection for Time Series Using VAE-LSTM Hybrid Model," in \textit{IEEE International Conference on Acoustics, Speech and Signal Processing (ICASSP)}, pp. 4322-4326, 2020.

\bibitem{deep5} A. Song, E. Seo, and H. Kim, ``Anomaly VAE-Transformer: A Deep Learning Approach for Anomaly Detection in Decentralized Finance," \textit{IEEE Access,}  vol. 11, pp. 98115-98131, 2023.

\bibitem{FAE} K. Elkhalil, A. Hasan, J. Ding, S. Farsiu, and V. Tarokh, ``Fisher Auto-Encoders," in \textit{International Conference on Artificial Intelligence and Statistics}, pp. 352-360, 2021.

\bibitem{prior} A. Kopf, V. Fortuin, V. R. Somnath, and M. Claassen, ``Mixture-of-experts Variational Autoencoder for Clustering and Generating from Similarity-based Representations on Single Cell Data," \textit{PLoS computational biology}, vol. 17, p. e1009086, 2021.

\bibitem{FAE2} G. Liu, L. Li, L. Jiao, Y. Dong,  and X. Li, ``Stacked Fisher Autoencoder for SAR Change Detection," \textit{Pattern Recognition}, vol. 96, p. 106971, 2019.


\bibitem{sensor3} T. S. Sowers, G. Kopasakis, and D. L. Simon, ``Application of the Systematic Sensor Selection Strategy for Turbofan Engine Diagnostics," \textit{Turbo Expo: Power for Land, Sea, and Air,} vol. 43123, pp. 135-143, 2008.
\bibitem{Fisher} J. Ding, R. Calderbank, and V. Tarokh, ``Gradient Information for Representation and Modeling," \textit{Advances in Neural Information Processing Systems,} vol. 32, 2019.

\bibitem{cmapss} A. Saxena, K. Goebel, D. Simon, and N. Eklund, ``Damage Propagation Modeling for Aircraft Engine Run-to-failure Simulation," \textit{ IEEE International Conference on Prognostics and Health Management}, pp. 1-9, 2008.


\end{thebibliography}
\end{document}